\documentclass[%
reprint,
superscriptaddress,
frontmatterverbose,
 amsmath,amssymb,
aps,
prb,
]{revtex4-2}

\usepackage{amsfonts,amsmath,amssymb,amsthm}

\usepackage{graphicx}
\usepackage{dcolumn}
\usepackage{bm,dsfont}

\usepackage{titlesec}

\usepackage{hyperref}

\hypersetup{
	pdfnewwindow=true,      
	colorlinks=true,       
	linkcolor=blue,          
	citecolor=blue,        
	filecolor=blue,      
	urlcolor=blue,          
}

\usepackage{color}

\usepackage{lineno,gensymb}
\makeatother

\newcommand{\be}{\begin{equation}}
\newcommand{\ee}{\end{equation}}

\def\ket#1{\mathinner{|{#1}\rangle}}

\newcommand{\re}[1] {\textcolor{red}{#1}}
\newcommand{\sro}{Sr$_4$RhO$_6$}
\newcommand{\jeh}{$J_{\text{eff}}$}

\usepackage{braket}
\usepackage{mathtools,siunitx}

\usepackage{changes}
\definechangesauthor[name={qqgu}, color=orange]{qg}

\begin{document}


\title{Emergence of bond-dependent highly anisotropic magnetic interactions in \sro: 
a theoretical study}

\author{Shishir Kumar Pandey}\email{shishir.kr.pandey@gmail.com}
\affiliation{International Center for Quantum Materials, School of Physics, Peking University, Beijing 100871, China}
\affiliation{AI for Science Institute, Beijing, China}


\author{Qiangqiang Gu}
\affiliation{AI for Science Institute, Beijing, China}
\affiliation{School of Mathematical Science, Peking University, Beijing 100871, China}

\author{Yihao Lin}
\affiliation{International Center for Quantum Materials, School of Physics, Peking University, Beijing 100871, China}

\author{Rajarshi Tiwari}
\affiliation{School of Physics, AMBER and CRANN Institute, Trinity College Dublin, Dublin 2, Ireland}

\author{Ji Feng}
\affiliation{International Center for Quantum Materials, School of Physics, Peking University, Beijing 100871, China}
\affiliation{Hefei National Laboratory, Hefei 230088, China}


%

\date{\today}

\begin{abstract}
The quantum spin liquid states as a natural ground state of the Kitaev model has led to a quest for new
materials candidates hosting Kitaev physics. Yet, there are very few material candidates in this category.
Using a combination of $ab$ $initio$ and model Hamiltonian methods, we propose that Ruddlesden-Popper compound \sro~belongs
to this category. With a tight-binding model and exact diagonalization approach, we show that
despite substantial trigonal-like distortion, the electronic and magnetic properties of \sro~can
be well described in terms of pseudo-spin = 1/2 states. Magnetic interactions among pseudo-spins,
estimated using the second-order perturbation method are highly bond-dependent anisotropic
in nature with two particularly noticeable features, antiferromagnetic Kitaev and  Dzyaloshinskii-Moriya interactions.
The gaped spin-wave spectra of \sro~obtained with linear spin-wave theory is consistent with
the underlying magnetic frustration. Additional analysis of the role of individual or a particular 
combination of magnetic interactions reveals that the spin-wave spectra of \sro~is a combined effect of 
the highly anisotropic interactions and a relatively simpler minimal model may not be plausible in the current 
case. The crucial insights about coupling between the local structural features and magnetic 
properties of \sro~obtained in this study may be helpful for future studies belonging to this class. 

\end{abstract}

\maketitle

\section{Introduction}
Orbital  and spin angular momentum of an electron are coupled through a relativistic effect called spin-orbit coupling (SOC).
Many interesting phenomena such as the anomalous Hall effect, manipulation of spin currents,
the emergence of topological properties in weakly correlated systems has been extensively studied~\cite{WTI_kane,WTI_hasan,WTI_szang} .
However, the strongly correlated materials provide host even richer physics because of the presence of additional interactions
such as crystal field splitting ($\Delta^{\text{\tiny CF}}$) and on-site Hubbard ($U$), often  competing with SOC~\cite{scorr_balents,SCorr_young}.
This competition gives rise to exotic phenomenon like realization of unconventional superconductivity~\cite{and,supercond1,supercond2},
emergence of topological phases~\cite{topo_nat_mat} and Kitaev physics~\cite{khul1}.
Among these examples, Kitaev physics~\cite{kitaev} particularly has recently got a lot of attention as a driving mechanism
in realization of quantum spin liquid states~\cite{qsl_mat_pres,nat_rev_qsl}.

The work of Jackeli and Khaliullin~\cite{khul1} accelerated the progress towards the realization of Kitaev physics in real materials.
Their proposal was based on magnetic interactions between pseudo-spins on a honeycomb lattice of transition metal ions originating
from the interplay of strong electrostatic crystal field (CF) of anions and SOC at transition metal sites.
The five degenerate $d$ orbitals of transition metal atom split into triply degenerate $t_{2g}$ and doubly degenerate $e_g$ orbitals
due to $\Delta^\text{\tiny CF}$ (see Fig.~\ref{fig:fig01}(a)).
Energetically lower $t_{2g}$ manifold further splits in the presence of SOC to form the half-filled pseudo-spin \jeh~$=\frac 12$ states dominating low energy space of materials. Magnetic interactions between these \jeh~$=\frac 12$ pseudo-spin states was proposed to be dominantly Kitaev-type. Cobaltates~\cite{co1,co2,co3,co4,co5,co6,co7,co8,co9,co_skp}, iridates~\cite{ir1, ir2, ir3,ir4,ir5,ir6,ir7, liiro_cf} and $\alpha$-RuCl$_3$~\cite{ru1,ru2,ru3,rucl3_bnds,rucl3_method} are some of the examples falling in this category. Recent studies
on Ir-based double
perovskite compounds have further widen the horizon of Kitaev physics on frustrated
fcc lattice formed by magnetic ions with spatially separated octahedral environment~\cite{dp1,dp2,dp3,dp4,dp5}.

These pseudo-spin \jeh~=$\frac 12$ doublets are Kramers's doublet which relate to each other by time-reversal symmetry and 
are degenerate when time-reversal symmetry is preserved.
The associated operators, \bm{\jeh$^\gamma$}, where $\gamma$ = $x$, $y$, $z$, thus follow the spin commutation relations.
Only in the
limits, $\Delta^\text{\tiny CF}$ $\to$ $\infty$ and when the splitting among the $t_{2g}$
manifold due to additional trigonal(tetragonal) distortions
$\Delta^\text{\tiny CF}_\text{\tiny tri.}$($\Delta^\text{\tiny CF}_\text{\tiny tet.}$) $\to$ 0, a
pure \jeh~=$\frac 12$ state can be realized.

However, the real materials mentioned above are far from these ideal limits making the situation even more complex.
Such complexities are inevitable when a minor change in details of these interactions may have dramatic effects on the macroscopic behavior of the material.
For example, in iridates despite the presence of additional
$\Delta^\text{\tiny CF}_\text{\tiny tri.}$($\Delta^\text{\tiny CF}_\text{\tiny tet.}$)
distortions which are responsible for mixing between \jeh~=$\frac 12$ and $\frac 32$ states
~\cite{ir4,liiro_cf}, the large SOC
of Ir 5$d$ orbitals still allow \jeh$=\frac 12$ description of the magnetic properties.
However, same cannot be pre-assumed for a 4$d$ transition metal compound where SOC strength is nearly
half of its 5$d$ counterpart and $\Delta^\text{\tiny CF}_\text{\tiny tri.}$($\Delta^\text{\tiny CF}_\text{\tiny tet.}$)
distortions of octahedra might be comparable to the SOC strength.
This inhibits any generic rule for behavior prediction of such materials and hence, a
case to case study is often required.

The scarcity of 4$d$ magnetic compounds with $J_{\text{eff}}=\frac 12$ behavior
makes it even more difficult to obtain any comprehensive understanding.
To the best of our knowledge, the only example of magnetic material in this category is $\alpha$-RuCl$_3$ and has been
the subject of extensive theoretical and 
experimental investigations~\cite{ru1,ru2,ru3,rucl3_bnds,rucl3_method}.
Other 4$d$ materials such as Li$_2$RhO$_3$, Sr$_2$RhO$_4$, and some theoretically predicted
Rh and Ir-based fluorides are either non-magnetic (Li$_2$RhO$_3$ shows spin-glass behavior)
or paramagnetic in nature~\cite{li2rho3,sr2rho4,rhirf}.
In the quest of new Kitaev candidates, \sro~is another possible example of a 4$d$ oxide~\cite{sro2,calder_sro}.
Materials like \sro~and some Ir-based double perovskites~\cite{dp1,dp2,dp3,dp4,dp5}) with
isolated metal-anion octahedra (as shown in Fig.~\ref{fig:fig01}(b)) may possess an advantage
over materials with edge-shared geometry because the larger spatial separation between the magnetic
ions in the former can minimize the direct overlap of $d$ orbitals as compared to edge shared geometry.
This in turn may result in suppression of additional $undesirable$ nearest-neighbor as well as farther
neighbor Heisenberg-like isotropic coupling.
\sro~is believed to exhibit
ideal cubic octahedral environment on Rh-sites~\cite{calder_sro} in a centrosymmetric crystal structure. Such a
distinctive feature may lead to the realization of
$pure$ \jeh~=1/2 states, a feature not realized in any of the previously mentioned Kitaev candidate materials.
Despite purportedly having such lucrative features with the possibility of hosting rich physics, it is surprising
to find no theoretical study dedicated to this material and hence is the focus of our study in this article.

In this study, using a combination of first-principles calculations and a tight-binding model,
we first show that contrary to the earlier belief~\cite{calder_sro}, the Rh-O$_6$ octahedra in \sro~is
not perfect and the octahedral crystal field at Rh-sites has additional trigonal-like distortions
originating from the influence of the extended environment of Sr atoms.
Using the exact diagonalization (ED) technique, we show that despite such a distortion,
mixing between $J_{\text{eff}}$ = 1/2-3/2 states is small and description of low-energy
space in terms of $J_{\text{eff}}$ = 1/2 states is still valid in this material.
Magnetic interaction among these pseudo-spins estimated using second-order perturbation theory
show highly bond-dependent anisotropic behavior with additional diagonal/off-diagonal terms appearing
alongside two particularly noticeable features, antiferromagnetic Kitaev and Dzyaloshinskii-Moriya (DMI)
interactions on some of the first nearest-neighbor (1NN) Rh-Rh bonds.
We attribute the appearance of DMI to the local inversion symmetry breaking due to the extended
environment of Sr$^{+2}$ ions.
The second and third nearest-neighbor interactions are found to be negligibly small.
The classically optimized magnetic ground state brings
an antiferromagnetic configuration which is energetically close to the previously proposed magnetic structure.
Spin wave spectra calculated using linear spin-wave theory is found to be
gaped throughout the Brillouin zone,
consistent with underlying frustrated magnetic frustration. Origin of various features of the spectra is
analyzed separately by examining the role of various magnetic interaction terms in the spin Hamiltonian.
This analysis establish the fact that the spectra is a combined effort of all these highly 
anisotropic magnetic interactions and a relatively simpler minimal magnetic model may not be plausible in the current
case.
Our study provides crucial insights for compounds belonging to this class.

 \section{Methods}

\subsection{{\it Ab initio} calculations} 

 Density-functional theory calculations have been performed using projector-augmented wave method \cite{PAW,PAWpotentials1}, implemented within Vienna $ab$ $initio$ simulation package (VASP)~\cite{Kresse}.
The Perdew-Burke-Ernzerhof functional \cite{PBE} is used for the exchange-correlation functional
within the GGA formalism.
We start with the experimental lattice parameters of trigonal crystal system of \sro~with centrosymmetric space group $R\bar{3}c$ (No. 167) which are $|${\bf a}$|$ = $|${\bf b}$|$ = 9.740 \AA{}, $|${\bf c}$|$ = 11.840 \AA{}; $\alpha$ = $\beta$ = 90\degree~and
 $\gamma$ = 120\degree~\cite{calder_sro}. Using plane wave cutoff energy 550 eV, $4\times4\times2$ $\Gamma$-centered $k$-mesh and energy convergence criteria of 10$^{-5}$ eV, we optimize the lattice parameters with experimentally proposed magnetic ground  
 (accommodated within 24 Rh atoms in a $2\times2\times1$ supercell) considering
 SOC effect at the self-consistent level. A DFT+$U$ approach employing Liechtenstein~\cite{liech} scheme with
 on-site Coulomb interaction $U$ = 2.5 eV and exchange interaction $J_\text{H}$ = 0.9 eV was used. The values of $U$ and
 $J_\text{H}$ parameters are consistent with the previous study~\cite{calder_sro}.
 Optimized $a$ and $b$ lattice constants were found
to be overestimated by $\sim$ 3.1 \% while $c$ remains the same. Since this change in lattice constants is significant,
we have used the optimized structure in further calculations.

\subsection{Estimation of electronic parameters}

Non spin polarised tight-binding (TB) Hamiltonian ($H_\text{TB}$) in local axes framework 
(see Fig.~\ref{fig:fig01}(b)) was calculated by projecting onto all the five
Rh-$d$ orbitals using the Wannierization procedure~\cite{wannier90} and
is shown in Fig.~\ref{fig:fig02}(a). 
On the two symmetry in-equivalent Rh sites, octahedron are rotated around the $C_3$-axis
which is along crystallographic \bm{$c$} axis. We choose the local axes 
(\bm{$x$}, \bm{$y$}, \bm{$z$}) along oxygen atoms 
on one of the Rh site obeying \bm{$c$} = \bm{$x$} + \bm{$y$} + \bm{$z$} and rotate these axes 
on the other Rh site by a unitary transformation to obtain the identical form of CF matrix 
on the two sites.
Crystal field matrix on a site $i$
($\Delta^\text{\tiny CF}_i$) is extracted
from the onsite part of $H_\text{TB}$ obeys crystals $C_3$ symmetry. 
To extract the SOC strength ($\lambda$), we fit the $ab$ $inito$ band structure,
where the SOC was included at the self-consistent level, with $H_\text{TB}$ after adding the onsite $H_{\text{soc}}$ = 
$\sum_i \lambda \bm L_i \cdot \bm s_i$ term~\cite{qqgu}. 
The fitting is shown in Fig.~\ref{fig:fig02}(C) with the inset showing the fitting near the Fermi level.
It brings $\lambda$ = 90 meV. This vaue is smaller than the considered value for iso-electronic 
$\alpha$-RuCl$_3$ ($\lambda$ = 140 meV) ~\cite{rucl3_method,rucl3_ref} and 
a recently estimated value of 175 meV for Rh atom~\cite{soc_est}. 
However, on a later stage, we will show that considering these three values does not bring any 
qualitative changes 
in the magnetic interactions and hence, for rest of the discussion in the manuscript we choose 
$\lambda$ = 140 meV. 
We estimate the Coulomb matrix elements $U_{ijkl}$($\omega$ = 0) within the constrained random phase approximation (cRPA).
To this end, we neglect the screening effects for all the five Rh $d$ orbitals states which are energetically well-separated form other states~\cite{crpa1,crpa2,crpa3}.  
The estimated parameters are
$U$ = 2.474 eV and $J_\text{H}$ = 0.106 eV which were further used in our multi-band Hubbard model.

\section{Results}
\subsection{Structural analysis and electronic properties}

\begin{figure}[ht]
\centering
\includegraphics[width=8.0 cm]{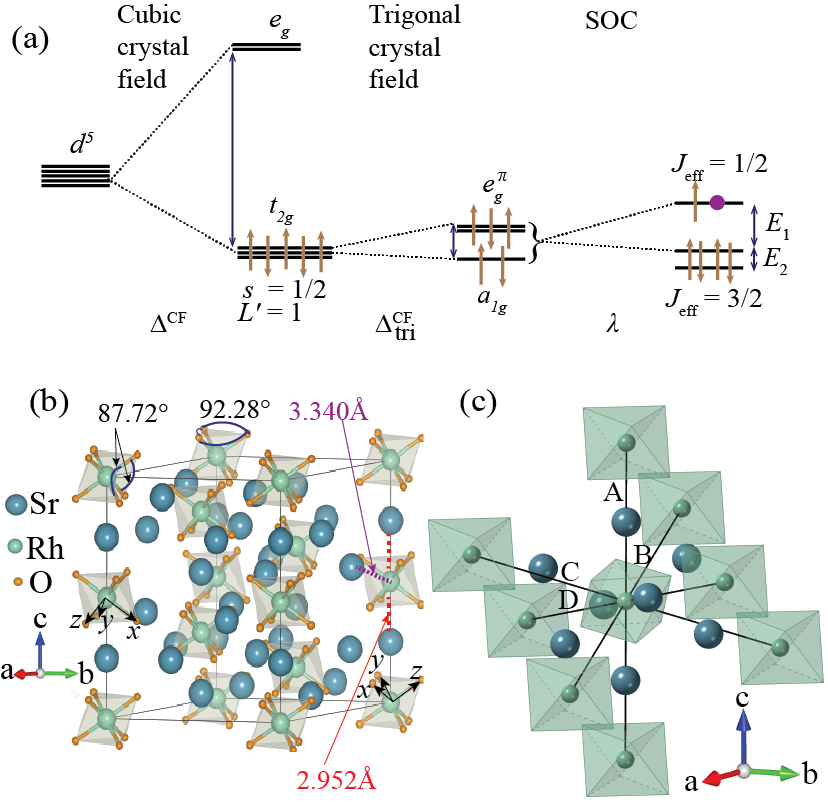}
\caption{(a) \jeh~picture for a $d^5$ system arising from octahedral crystal field ($\Delta^\text{\tiny CF}$) and spin-orbit coupling ($\lambda$).
Additional splitting of  $t_{2g}$ states into a singlet $a_{1g}$ and a doublet $e_g^\pi$ 
due to trigonal like distortions ($\Delta^\text{\tiny
CF}_\text{\tiny tri.}$). SOC further leads to a \jeh~=1/2 doublet and two \jeh~=3/2 doublets separated by $E_2$. 
$E_1$ is the energy separation between \jeh~=1/2 and closest \jeh~=3/2 doublet.
(b) Side view of \sro~crystal structure. Spatially separated octahedron are evident. Local $x$, $y$, and $z$ axes on two of the
octahedron are shown. $a$, $b$ and $c$ are the global crystallographic axes. Two kinds of color-coded Rh-Sr bonds along with
O-Rh-O bond angles obtained after optimization of the crystal structure are shown.
(c) Extended local environment (including Sr atoms) around an Rh atom in \sro.
Four types of Rh-Rh nearest neighbors A, B, C, and D-bonds with Rh-O Octahedra on these bonds are also shown.}
  \label{fig:fig01}
\end{figure}

 Under a large $\Delta^\text{\tiny CF}_i$, the low energy space Rh-4$d^5$ ions can be described by a single hole within the $t_{2g}$ manifold with effective spin moment \textit{s}  = 1/2 and effective orbital angular moment \textit{L$'$} = 1. The spin-orbit coupling then leads to  an effective total angular momentum of \textbf{\textit{J}$_\text{eff}$} = \textbf{\textit{s}} - \textbf{\textit{L$'$}} resulting in doubly degenerate
   pseudospin-1/2 states forming low-energy space in this material.
   This is schematically shown in Fig.~\ref{fig:fig01}(a). However, the lowering of cubic $O_h$ symmetry of octahedron
   due additional $\Delta^\text{\tiny CF}_\text{ \tiny tri.}$($\Delta^\text{\tiny CF}_\text{\tiny tet.}$) terms can
   invalidate this picture. Hence it is important to first examine whether Rh-O$_6$ octahedra in \sro~retains the $O_h$ symmetry as was
   proposed earlier in Ref~\cite{calder_sro}.
   
In this experimental crystal structure, all the six Rh-O bond lengths are $\sim$ 2.044 \AA{} while O-Rh-O bond angles are quite
close to the ideal 90\degree{} with the largest deviation being 0.1\degree. However,
full structural optimization with the magnetic ground state in our DFT calculation brings
substantial changes in $a$ and $b$ lattice constants along with the changes in the local octahedral
environment. The optimization enhanced $a$ and $b$ lattice constants to 
10.046 \AA{} and also 
all the six Rh-O bond lengths elongated to 2.109 \AA{}. The structural optimization also alters
the O-Rh-O bond angles to 92.28 and 87.72\degree (see Fig.~\ref{fig:fig01}(b)).
Also, out of eight Sr neighbors in
the extended environment of Rh atoms, two ``apical'' Sr atoms along
$c$ axis are at 2.952 \AA{} distance while other ``non-apical'' six are at 3.340 \AA{} in the optimized structure. This is shown in  
Fig.~\ref{fig:fig01}(b) (short Rh-Sr distances are along A-bonds and long ones are along B/C/D-bonds).
These two kinds of Rh-Sr distances were 2.960 and 3.238 \AA{} in the starting structure.
Almost similar Rh-Rh 1NN distances $\sim$ 5.98/6.0 \AA{} of all the 8 bonds before optimization has changed substantially now
to $\sim$ 5.9 and 6.13 \AA{} for
two A-bonds, and six B/C/D bonds respectively.
Thus optimization of structure results in substantial changes in Rh-O, non-apical Rh-Sr, and overall Rh-Rh bond lengths and is
 consistent with enhancement of $a$ and $b$ lattice constants.
 
\begin{figure}[ht]
\centering
\includegraphics[width=7.0 cm]{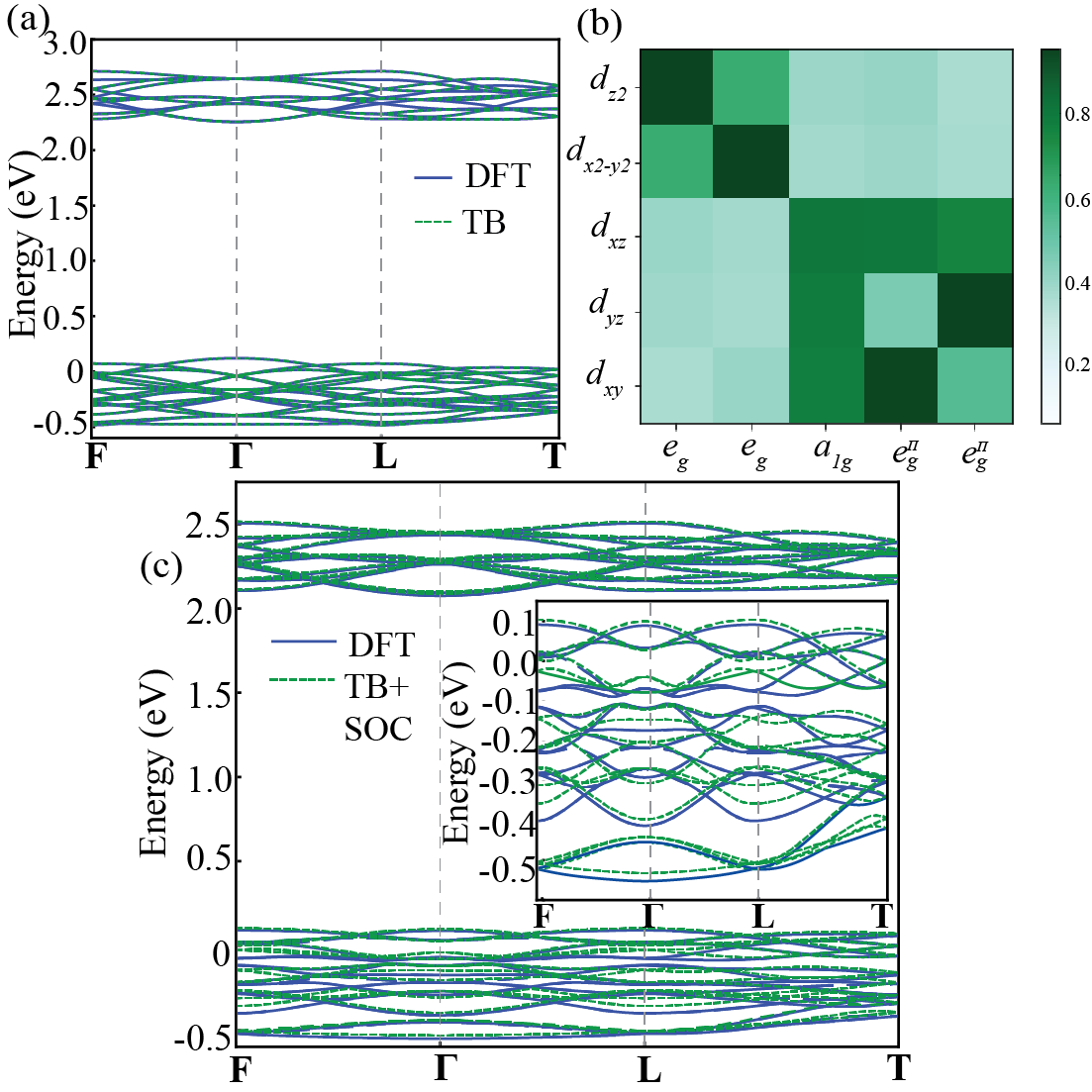}
\caption{(a) Band structure plot from the $ab$ $initio$ and Wannier-based 
TB model calculations considering all the Rh $d$ orbitals in the basis. 
(b) Graphical representation of  
eigenvectors of matrix in Eq.~\ref{eq:cf}. In column-wise representation of eigenvectors each row represents
absolute weight of individual orbitals. Labeling of eigenstates is done in accordance of Fig.~\ref{fig:fig01}(a).
(c) Fitting of $ab$ $initio$ SOC band structure  with Wannier based tight binding model after including onsite SOC term 
in the 
Hamiltonian. The inset shows the fitting near the Fermi level which is set to zero in all the plots.}
  \label{fig:fig02}
\end{figure}

In order to understand how these changes in crystal structure affect the CF, we set up a TB model with 
$d$ orbital basis of $\psi^{\dagger}$ = $[d^\dagger_{z^2}$, $d^\dagger_{x^2-y^2}$, $d^\dagger_{xz}$,
$d^\dagger_{yz}$, $d^\dagger_{xy}]$ using Wannierization procedure as mentioned in Methods section
(fitting is shown in Fig.~\ref{fig:fig02}(a)).
CF matrix $\Delta^\text{\tiny CF}_i$ obtained from $H_\text{TB}$ 
is given in Eq.~\ref{eq:cf}. Entries in the matrix are in the unit of eV. 
One can clear see that this CF matrix obeys $C_3$ symmetry restriction as the off-diagonal elements 
within the $t_{2g}$  (colored entries in the matrix) manifold have nearly the same absolute values 
within an error bar of 5 meV.

\begin{equation}
\label{eq:cf}
\Delta^\text{CF}_i =\begin{bmatrix*}[r]
   2.5379 &  -0.0006  &  -0.1292   &  0.0987  &  -0.0517 \\
  -0.0006 &   2.5474  &   0.0150   & -0.0088  &  -0.0762 \\
  -0.1292 &   0.0150  &   0.1127   &  \textcolor{brown}{0.0499}  &   \textcolor{brown}{0.0531} \\
   0.0987 &  -0.0088  &   \textcolor{brown}{0.0499}   &  0.1055  &  \textcolor{brown}{-0.0547} \\
  -0.0517 &  -0.0762  &   \textcolor{brown}{0.0531}   & \textcolor{brown}{-0.0547}  &   0.1138 \\
\end{bmatrix*}
\end{equation}

By diagonalizing this matrix one can find 
that the $t_{2g}$-$e_g$ crystal field splitting ($\Delta^{t_{2g}-e_g}_i$) 
is $\sim$ 2.630 eV while 
the triply degenerate $t_{2g}$ splits into a $a_{1g}$ singlet  and $e_g^\pi$ doublet by 
$\Delta^\text{\tiny CF}_\text{\tiny tri}$ $\sim$ 160 meV with doublet being higher in energy than the singlet.
The corresponding eigenvectors are graphically represented in Fig.~\ref{fig:fig02}(b), where each column in the 5 $\times$ 5 
graph represents an eigenvector with row representing absolute weight of individual orbitals. This representation 
clearly highlights the nature of $\Delta^\text{\tiny CF}_i$ in \sro~which has been depicted in Fig.~\ref{fig:fig01}(a). 

This particular form of $\Delta^\text{\tiny CF}_i$ can be understood as follows.
The shorter ``apical'' Sr-Rh bond passes through the center of two
triangular faces of Rh-O$_6$ octahedra
as shown in Fig.~\ref{fig:fig01}(b)-(c). This bond is
 also one of the four three-fold rotational symmetry ($C_3$) axis of the Rh-O$_6$ octahedra.
 The electrostatic repulsion along these shorter bonds behaves as compressing strain causing
 changes in Rh-O bond lengths and O-Rh-O bond angles.
 This is analogous to the case of trigonal distortions where
 bond distortions take place along one of the four $C_3$ axes of the octahedra.
Thus, in \sro, an extended anisotropic environment of Sr atoms produces a non-spherical crystalline potential
responsible for additional $\Delta^\text{\tiny CF}_\text{\tiny tri.}$ of Rh-O$_6$ octahedra.
The cubic $O_h$ symmetry, then lowers to $C_{3i}$ (-3) in this case. 

\begin{figure*}[ht]
\centering
\includegraphics[width=16.0 cm]{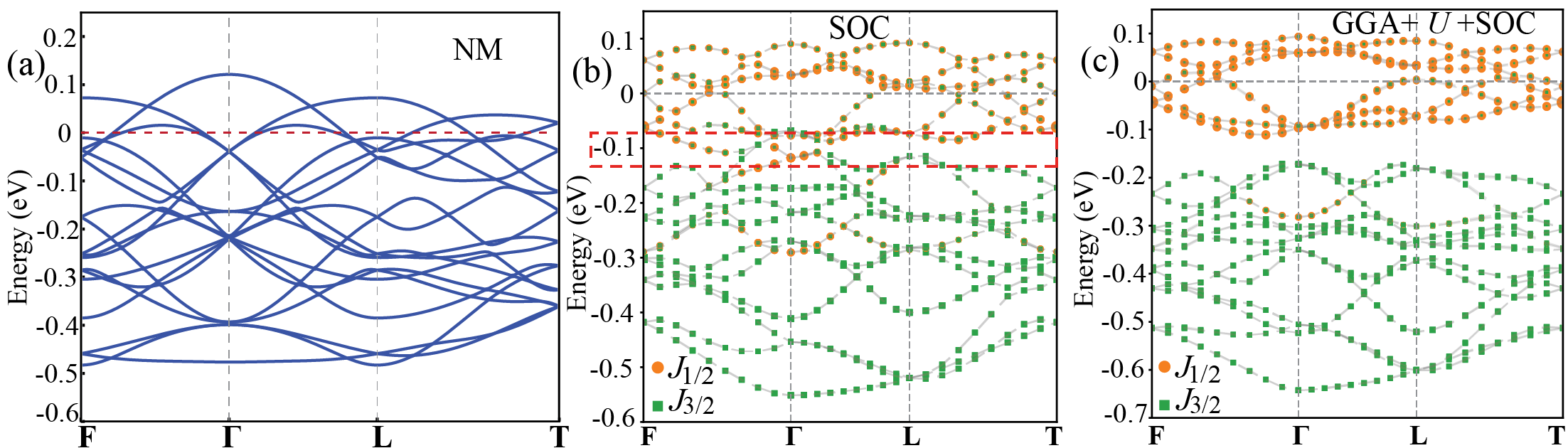}
\caption{\textcolor{red}{$Ab$ $initio$ band structure plots near the Fermi level. Case of (a) non-magnetic
(NM), (b) SOC included and (c) SOC + $U$ (on Rh $d$ orbitals) band structure projected onto \jeh~states. 
Fermi level is set to 0 eV. The fined dashed box in (b) shows the energy window of  
 where separation of  \jeh~=1/2 bands from other \jeh~=3/2 bands take place with
 inclusion of SOC. Clear \jeh~= 1/2 character is apparent at the Fermi level in (c).}}
  \label{fig:fig03}
\end{figure*}

 Distortions like $\Delta^\text{\tiny CF}_\text{\tiny tri.}$ tend to lower the energy separation
 between \jeh~=1/2 and 3/2 states. This brings a genuine concern about effect of SOC on electronic
 structure of \sro~and whether the strength of SOC in \sro~is
 sufficient enough to separate out these two states.
  To examine this point, we calculated
the $ab$ $initio$ band structures for three cases, (i) non-magnetic, (ii) with SOC, and (iii) with
 SOC + $U$. SOC was included at the self-consistent level in these calculations.
 
 In Fig.~\ref{fig:fig03}, we only show the bands near the Fermi level which are dominantly contributed by
 $t_{2g}$ orbitals. \textcolor{red}{We projected the band structures onto $J_\text{eff}$ states with the form given below. 
\begin{eqnarray} \nonumber
  \Ket{\frac{1}{2}, \pm \frac{1}{2}} &=& \frac{1}{\sqrt{3}}\left( \mp \ket{d_{xy}, \pm \frac{1}{2}} \mp i \ket{d_{xz}, \mp \frac{1}{2}} - \ket{d_{yz}, \mp \frac{1}{2}} \right) \\ \nonumber
  \Ket{\frac{3}{2}, \pm \frac{3}{2}} &=& \frac{1}{\sqrt{2}}\left( -i \ket{d_{xz}, \pm \frac{1}{2}} \mp \ket{d_{yz}, \pm \frac{1}{2}}  \right) \\ \nonumber
  \Ket{\frac{3}{2}, \pm \frac{1}{2}} &=& \frac{1}{\sqrt{6}}\left( 2 \ket{d_{xy}, \pm \frac{1}{2}} -i\ket{d_{xz}, \mp \frac{1}{2}} \mp \ket{d_{yz}, \mp \frac{1}{2}} \right) \nonumber
\end{eqnarray}
There are mainly two points to be noticed in Fig.~\ref{fig:fig03}. First that inclusion of SOC substantial changes the band structure. This is apparent from 
comparing non-spin polarised band structure plot in Fig.~\ref{fig:fig03}(a) and SOC included band structure plot shown in Fig.~\ref{fig:fig03}(b).
In particular, SOC leads to separation of \jeh~= 1/2 bands near -0.1 eV (red box in  Fig.~\ref{fig:fig03}(b)) from the other bands (\jeh~= 3/2 bands)  near -0.1 eV.
Inclusion of $U$ on Rh $d$ states further contributes to this band separation as shown in Fig.~\ref{fig:fig03}(c) and the dominant contribution near the Fermi level now 
clearly shown to have \jeh~= 1/2 character.}
Imposition of the magnetic ground state in band structure calculation (not shown) fully opens the gap at the Fermi level
 making it insulating.
 This is similar to the case of
 $\alpha$-RuCl$_3$~\cite{rucl3_bnds}.
From this analysis of electronic structure, one can conclude that the electronic
structure of \sro~is the combined efforts of $U$, SOC, and magnetism. Having examined the role of SOC,
one can further quantify the \jeh~=1/2 and 3/2 states admixture due to $\Delta^\text{\tiny CF}_\text{\tiny tri.}$ by considering
a multi-band Hubbard model for an isolated Rh$^{+4}$ ion. This is discussed in the next section.
 



\subsection{Onsite Hamiltonian and the atomic features}

One way to estimate the extent of mixing between the \jeh~=1/2 and 3/2 states is by calculating the
projection of a ``pure'' \jeh~=1/2 and 3/2 states for the case when
$\Delta^\text{\tiny CF}_\text{\tiny tri.}$ = 0
onto the ``true''  \jeh~=1/2 obtained with Eq.~\ref{eq:cf}. These states are
atomic features and hence can be described in an isolated atom limit. In this limit, a multiband Hubbard Hamiltonian
 at site, $i$ in the  five-orbital basis reads as,
\begin{eqnarray}
H_0 &=& H_{\text{cf}} + H_{\text{soc}} + H_{\text{int}} \nonumber \\
&=& \sum_{i,\sigma} \psi_{i\sigma}^\dagger {\Delta^\text{CF}_i}\psi_{i\sigma} + \sum_i \lambda \bm L_i \cdot \bm s_i \nonumber \\ 
&+& \frac{U}{2} \displaystyle\sum_{i,\alpha} n_{i\alpha \sigma} n_{i\alpha \sigma'} 
+ \frac{U'}{2} \displaystyle\sum_{i,\alpha \ne \beta} n_{i \alpha}n_{i \beta} \nonumber\\
 \nonumber &-&\frac{J_\text{H}}{2} \displaystyle\sum_{i, \sigma, \sigma', \alpha \ne \beta} \psi^\dagger_{i\alpha \sigma} \psi_{i\alpha \sigma'} \psi^\dagger_{i\beta \sigma'} \psi_{i\beta \sigma} \\ 
 &-& \frac{J'}{2} \displaystyle\sum_{i, \sigma \ne \sigma', \alpha \ne \beta} \psi^\dagger_{i\alpha \sigma} \psi_{i\beta \sigma'} \psi^\dagger_{i\alpha \sigma'} \psi_{i\beta \sigma} 
\label{eq:h0}
\end{eqnarray}
 In above expression,  $U$/$U'$ are
intraorbital/interorbital Hartree energies;  and  $J_\text{H}$ and $J'$ are Hund's coupling
and pair hopping interaction,
respectively. Rotational invariance in the isolated atom limit dictates the relationships:  $U'$ = $U$ - 2$J_\text{H}$ and $J_{\text H}$ = $J'$.
We use $U$ = 2.474 eV and $J_\text{H}$ = 0.106 eV  which are
estimated from cRPA as mentioned in the Methods section and $\lambda$ = 140 meV is considered. 
We diagonalize the above Hamiltonian considering five electrons of Rh$^{+4}$
ions which give a total of 252 eigenstates, the lowest two and the next four of which are the \jeh~=1/2 states and \jeh~=3/2 states, respectively.

\begin{table}[h!]
 \centering
\begin{tabular}{llll}
\hline{}
   &\multicolumn{2}{c}{$\braket{\phi'_{\alpha}|\phi_{\beta}}$}  \\
\cline{2-3}
 $\bra{\phi'_{\alpha}}$   & $\ket{\phi_{1}}$ & $\ket{\phi_{2}}$ \\
\hline
 1 &  0.901     &   0.352     \\
 2 & 0.352      &  0.901      \\
3 & 0.162      & 0.056 \\
4 & 0.056        & 0.162 \\
5 & 0.101        & 0.117 \\
6 & 0.117        & 0.101 \\
\hline 
\end{tabular}
\caption{Projections of \jeh~=1/2, 3/2 states obtained when $\Delta^\text{\tiny CF}_\text{\tiny tri.}$ = 0,  onto \jeh~=1/2 states
with $true$ CF from Eq.~\ref{eq:cf}. These states are obtained from exact-diagonalization of the Hamiltonian in Eq.~\ref{eq:h0}.} 
\label{tab:01}
\end{table}

For $\Delta^\text{\tiny CF}_\text{\tiny tri.}$ = 0,  $t_{2g}$-$e_g$ splitting was fixed at 2.790 eV 
and all the off-diagonal matrix elements were zeroed in Eq.~\ref{eq:cf}. 
The lowest six states in this case are represented by 
 $\{\phi'_{\alpha}\}$, $\alpha$ = 1, 6  while lowest two states obtained using $true$ CF from Eq.~\ref{eq:cf} are 
 labelled as 
 $\{\phi_{\beta}\}$,  $\beta$ = 1-2. The projections $\braket{\phi'_{\alpha}|\phi_{\beta}}$ 
are listed in Table~\ref{tab:01}.
From the table, since $|\braket{\phi'_{\alpha}|\phi_{\beta}}|^2$ = 0.811
for $\delta_{\alpha\beta}$ = 1, 2, one can conclude that the \jeh~=1/2 
states retain their major weight despite a substantial 
$\Delta^\text{\tiny CF}_\text{\tiny tri.}$, validating applicability of 
\jeh~= 1/2 picture in \sro. The non-zero value of projections 
$|\braket{\phi'_{\alpha}|\phi_{\beta}}|^2$ ($\sim$ 0.026/0.010) for $
\alpha$ = 3-6, $\beta$ = 1-2 indicates a small admixture  of  \jeh~=1/2 and 3/2 states due to $\Delta^\text{\tiny CF}_\text{\tiny tri.}$. We find small changes of $\sim$ 4 \% in these projections 
for $\lambda$ = 90 meV.

One of the quantities which can be measured from the resonant
inelastic X-ray scattering experiments are the single-point excitations represented by sharp peaks in the
scattering intensity in the relevant energy range.  
It can be a direct probe for cubic symmetry lowering of the Rh-O$_6$ octahedra in \sro.
Theoretically, such a low-lying crystal field-assisted many-body
excitations bear a  
close resemblance to the eigenvalues obtained from diagonalization of many-body
Hamiltonian in Eq.~\ref{eq:h0}. For \sro, analysis of eigenvalue reveals that
 the \jeh~= 3/2 states split into two doublets by $E_2$ = 0.133 eV (see Fig.~\ref{fig:fig01}(a)) which would
otherwise be four-fold degenerate if $\Delta^\text{\tiny CF}_\text{\tiny tri.}$ = 0.
Energy separation of the \jeh~= 1/2 doublet with the lower \jeh~= 3/2 doublet is $E_1$ = 0.181 eV.
It can also be observed that $E_1$ is $\sim$ 30 meV
smaller than the expected value of   
$\frac{3}{2}\lambda$ due to finite $\Delta^\text{\tiny CF}_\text{\tiny tri.}$.  
From the higher \jeh~= 3/2 doublets, the next single ion excitation
is at $\sim$ 1.695 eV. From this point, a broad continuum of states with energy separations of
few meV in the window of  
$\sim$ 165 meV are found in our calculations. Having investigated the electronic properties of
\sro, we now discuss its magnetic properties in the next section.

\subsection{Magnetism}

 We start by projecting the 
Hamiltonian in Eq.~\ref{eq:h0} to the pseudo-spins $J_{1/2}$ subspace and introduce hoping 
($H_\text{hop}$) as perturbation. The hopping amplitudes are extracted from $H_\text{TB}$ 
and are listed in Appendix~\ref{appendix:hop}.
In the limit $U \gg t$, the second-order perturbation term brings, 
\begin{widetext}
\begin{eqnarray}
H^{(2)} &=& \sum_{ij}\sum_{\alpha\beta\alpha'\beta'} \mathcal H (i,j)_{\alpha\beta\alpha'\beta'}
| i\alpha,j\beta\rangle\, \langle i\alpha', j\beta '|, \nonumber\\
\mathcal H (i,j)_{\alpha\beta\alpha'\beta'} &=&
\sum_{kl}\sum_{\gamma\lambda}
\frac{1 }{\Delta E}
\langle i\alpha,j\beta|H_{\text{hop}}|k\gamma, l\lambda\rangle\,
\langle k\gamma,l\lambda|H_{\text{hop}}|i\alpha', j\beta'\rangle ,
\label{eq:h02}
\end{eqnarray}
\end{widetext}
where 
$1/\Delta E = \frac{1}{2}[{1}/(E_{i\alpha}+E_{j\beta} - E_{k\lambda}-E_{l\gamma})$.
Here, $| i\alpha,j\beta\rangle$ and $| i\alpha', j\beta'\rangle$ are two-site states made of  $J_{1/2}$ doublets, and $| k\lambda, l\gamma\rangle$ are two-site excited states with 
$d^6$ and $d^4$ configurations with Hilbert space dimensions of 210
for both.  
$H_{\text{hop}}$ connects a two-site ground state to these excited states.   
The eigenstates of isolated Rh ions with 4 and 6 -$d$ electrons are obtained again by exact diagonalization.

One can represnt the pseudo-spins $J_{1/2}$ as $S^{\mu} = \braket{i\alpha| \bm{J^\mu_{i,\text{ \bf eff}}}|i\beta}$
which are the expectation values of pseudospin {\bf $J^\mu_\text{eff}$} operators with $\mu = 0,x,y,z$. 
Here, \bm{$J^0_\text{\bf eff}$} = $\mathds{1}_{2\times2}$ is the matrix representation of operator \bm{$J^0_\text{\bf eff}$}. 
Using it, Eq.~(\ref{eq:h02}) can be mapped to a spin Hamiltonian of the form,
\begin{eqnarray}
H_\text{spin} &=& S_i^\mu \Gamma(i,j)^{\mu \nu}S_j^\nu \nonumber \\ 
&=& \Gamma(i,j)^{\mu\nu}  \phi _{i\alpha }^{\dagger }S _{\alpha\alpha '}^\mu \phi _{i \alpha '} 
 \phi _{j\beta } S_{\beta\beta '}^{\nu}  \phi _{j\beta }^{\dagger }, \nonumber
\end{eqnarray}
In the above expression, summation over all repeated indexes is implied. 
The map can be achieved by solving the linear equations,
\begin{equation}
-S_{\alpha\alpha'}^\mu S_{\beta\beta'}^\nu 
\Gamma(i,j)^{\mu\nu}
=\mathcal H(i,j)_{\alpha\beta\alpha'\beta'} \nonumber
\end{equation}

Here, degeneracy of the Kramers doublet leads $\Gamma^{0\mu}$ = $\Gamma^{\mu 0}$ = 0.
Thus, the most general form of exchange interaction matrix on an Rh-Rh 
bond $l \in$ $(i,j)$ is defined as, 
\begin{equation}
\label{eqn:jk}
    \Gamma_l = 
\begin{pmatrix}
J + \zeta & \eta + D & \eta' - D' \\
\eta - D & J -  \zeta & \eta'' + D'' \\
\eta' + D' & \eta''- D'' & J + K
\end{pmatrix}
\end{equation}

In the above expression, $J$, $K$, and $\eta$/$\eta'$ are the Heisenberg,  
Kitaev and off-diagonal interaction terms between the pseudospins-1/2,
while $\zeta$ is the diagonal anisotropic term.  DMI is represented
by ($D$, $D'$, $D''$) vector.

\begin{center}
\begin{table*}[ht]
 \centering
\begin{tabular}{lSSSSSSSSSSSSSS}
\hline
& & \multicolumn{3}{c}{$\lambda$ = 90 meV} & & & \multicolumn{3}{c}{$\lambda$ = 140 meV} & & &  \multicolumn{3}{c}{$\lambda$ = 174 meV} \\
 \cline{3-6} \cline{8-11} \cline{13-15}
  Term & &  A & B & C & & &  A & B & C & & & A & B & C\\ 
  \hline
  $J$      & &  0.149 & -0.519  &  4.262 & & &  0.301 & -0.109  &  3.250 & & &  0.402 & -0.021  &  2.975 \\ 
  $K$      & &  0.010 & -1.737  &  0.473 & & &  0.015 & -1.596  &  0.257 & & &  0.017 & -1.544  &  0.193 \\ 
  $\zeta$  & & -0.016 & -0.488  & -0.199 & & & -0.017 & -0.538  & -0.125 & & & -0.017 & -0.555  & -0.101 \\
  $\eta$   & &  0.022 & -2.246  &  1.066 & & &  0.017 & -1.829  &  0.460 & & &  0.011 & -1.686  &  0.296 \\
  $\eta'$  & & -0.016 &  1.040  & -0.873 & & & -0.014 &  0.683  & -0.348 & & &  0.000 &  0.564  & -0.207 \\ 
  $\eta''$ & &  0.025 & -1.231  &  0.723 & & &  0.019 & -0.899  &  0.271 & & &  0.012 & -0.784  &  0.154 \\
  $D$      & &  0.000 & -1.377  &  0.000 & & &  0.000 & -0.666  &  0.000 & & &  0.000 & -0.472  &  0.000 \\
  $D'$     & &  0.000 &  2.518  &  0.000 & & &  0.000 &  1.608  &  0.000 & & &  0.000 &  1.332  &  0.000\\
  $D''$    & &  0.000 & -2.303  &  0.000 & & &  0.000 & -1.325  &  0.000 & & &  0.000 & -1.041  &  0.000 \\
\hline 
\end{tabular}
\caption{ Estimated first neighbor (NN) Heisenberg $J$, Kitaev $K$ and diagonal $\zeta$  and off-diagonal $\eta$, $\eta'$, $\eta''$ anisotropic terms for \sro~given in meV. The second nearest neighbor interactions were found to be negligibly small ($<$ 0.01 meV). Parameters used are $U$ = 2.474 eV, 
$J_{\text{H}}$ = 0.106 eV and three values of $\lambda$ = 90, 140, 174 meV.}
\label{tab:gtable}
\end{table*}
\end{center}

The Rh atoms forms a body-centered cubic lattice in \sro~and thus
each Rh atom has eight 1NNs. Based on the nature of magnetic
interactions between different 1NNs, we subdivide the Rh-Rh bonds
 into three distinct categories which are indicated as  A/B/C/D bonds in Fig.~\ref{fig:fig01}(c).
Values of magnetic interactions are listed in Table~\ref{tab:gtable}.
For bond A and C, the $\Gamma_l$ matrix acquire a more symmetric form since on these bonds $\zeta$ = $\eta$ = $\eta'$
 = $\eta''$ = $D$ = $D'$ =$D''$ = 0.
 However, the magnetic interactions
on these two bonds differ in their strengths.
On B-bond,  $\Gamma_\text{B}$ takes the general form of Eq.~\ref{eqn:jk} and
$\Gamma_\text{D}$ can be obtained by simply taking the transpose of $\Gamma_\text{B}$.

Several remarks are in order.
First, one can see that the strength, as well as signs of interactions, differ for different bonds.
For example, for A and C bonds $J$, $\eta$, and $\eta'$ are antiferromagnetic while for B-bond they are ferromagnetic and the antiferro
Kitaev coupling is stronger on B-bond than the others. We emphasize that the antiferromagnetic Kitaev coupling in \sro,
although smaller, distinctly differs from the previous reports on iridates and $\alpha$-RuCl$_3$~\cite{rucl3_method,rucl3_ref}.
Second, quite interestingly,  DMI appears on B and D bonds in the centrosymmetric structure of \sro.
However, $D$, $D'$, and $D''$ have opposite signs on these two bonds. 
We attribute appearance of DMI to the local inversion symmetry
breaking due to
anisotropic crystalline potential produced by Sr atoms in the extended environment around Rh atoms 
shown in Fig.~\ref{fig:fig01}(c). The hopping pathways for the first nearest symmetry in-equivalent 
Rh-Rh neighbors gets influenced by the crystalline potential produced by this extended environment 
resulting in $T_{ij}^t \ne T_{ij}$ form of hopping matrix in appendix~\ref{appendix:hop}. 
Disappearance of DMI on A and C bond is merely an artifact of local coordinate system that we choose 
for our $H_\text{TB}$. For DMI between two sites, it is always possible to make a local rotation of the
spin coordinate axes at one of the sites to ``gauge" away this interaction by 
rotating the coordinates around the axis of the DM-vector by an angle corresponding to
the classical canting angle~\cite{dmi}. We verify this point by choosing a set of different local
axes in which DMI appears at both A and C bonds albeit smaller than B and D bonds.
Third, one may think that the Sr$^{+2}$ ions on A-bond may mediate superexchange interaction between Rh atoms
through their $s$ orbitals. However, on the contrary, we find highly suppressed interactions on this bond
suggesting a destructive role of the anisotropic crystalline potential of the Sr$^{+2}$ on magnetic interactions. 
Fourth, we found large off-diagonal terms on some of the Rh-Rh bonds. This is similar to the case of
iridates and $\alpha$-RuCl$_3$~\cite{rucl3_ref} resulting from  substantial $\Delta^\text{\tiny CF}_\text{\tiny tri.}$ distortions present
in all these materials. Based on the two particularly noticeable features in the first two points, $viz$-$a$-$viz$ 
antiferromagnetic Kitaev terms and appearance of DMI, one may consider \sro~a distinct 4$d$ magnetic material.

\begin{figure}[ht]
\centering
\includegraphics[height = 7.0 cm, width = 6.0 cm]{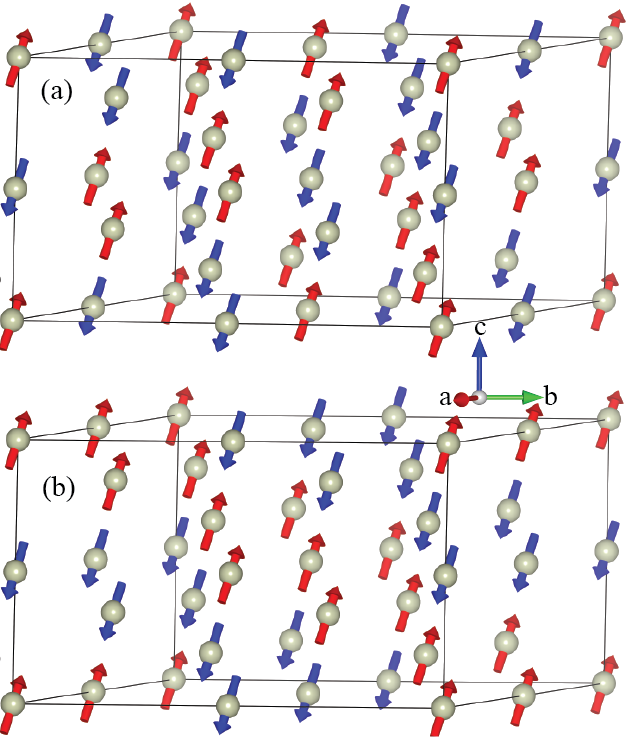}
\caption{(a) Experimentally proposed magnetic ground state of \sro.
(b) Classical ground state obtained from optimization of classical ground state using exchange interactions
of Table~\ref{tab:gtable}. Color coded spin orientation of only Rh lattice is shown here.}
  \label{fig:fig04}
\end{figure}

Varying the magnitude of SOC strength $\lambda$ in our model does not
change the interactions at a qualitative level. Estimated magnetic interactions for 
$\lambda$ = 90, 195 meV are listed in Table~\ref{tab:gtable} along with values for $\lambda$ = 140 meV.
The trend here is that with increase of $\lambda$, absolute values of all the magnetic interactions 
decreases except the AFM $J$ term on A bond. 

Magnetic interaction of Table~\ref{tab:gtable} are used to optimize the classical
magnetic state using SpinW package~\cite{spinw}.
The obtained magnetic ground state, represented by ordering vector $\sim$(1.0 0.5 0),
is shown in Fig.~\ref{fig:fig04}(b) along with the
experimentally proposed one in Fig.~\ref{fig:fig04}(a). The antiferromagnetic state obtained in our
calculations successfully captures most of the experimental features.
In the experimental magnetic structure,
the spin arrangement on Rh-Rh bonds(Fig.~\ref{fig:fig01}(c)), A and B are antiferromagnetic while on C and D
it is ferromagnetic. Optimized magnetic state in Fig.~\ref{fig:fig04}(b) from our calculations retains
antiferromagnetic coupling on A and ferromagnetic coupling at D bonds. However, this configuration differs from
the one shown in Fig.~\ref{fig:fig04}(a) on bonds B and C where the spin arrangement in the
two cases are just opposite to each other i.e. on B-bond the coupling is ferro
while on C-bond it is antiferromagnetic in our optimized structure.
Swapping the interactions at bonds B and C does
not bring the experimentally observed ground state indicating
 a joint meticulous effort of all the magnetic interactions to bring the ground state.
We find a slight deviation of magnetic moments from the $ac$ plane mainly due to the presence of
off-diagonal terms like $\eta$/$\eta'$/$\eta''$ and DMI. This is consistent with the
experimental finding of small tilting from $c$ axis~\cite{calder_sro}.
Our optimized magnetic configuration is energetically close to the experimentally proposed one with the former stabilized by
1.552 meV/spin.
The second and third neighbor magnetic interactions
are found to be negligibly small in \sro~and do not bring any
distinguishable change in the
optimization of the magnetic ground state. Thus we ignore them in further calculations of spin-wave spectra.

Here, we would like to comment that the scale of magnetic ordering temperature
of a material depends on various parameters like the strength of exchange interactions, number
of neighbors, their corresponding exchange contributions, and spatial dimensions of the magnetic lattice.
Although, the magnetic lattice of Rh atoms in \sro~form a three-dimensional bulk structure with eight
first magnetic neighbors,
the strongly frustrated anisotropic nature of bond-dependent magnetic interactions might be the reason
behind its experimentally observed low $T_\text{N}$ of $\sim$ 7.5 K. \re{We used the classical Monte Carlo technique implemented in
SpinW package~\cite{spinw} to estimate $T_\text{N}$ for \sro. Estimated value $T_\text{N}$  = 10.5 K 
 for interactions corresponding to $\lambda$ = 140 meV in Table~\ref{tab:gtable} is in close agreement with the experimental observation.}

\subsection{Spin-wave spectra}

\begin{figure}[ht]
\centering
\includegraphics[width=6.0  cm]{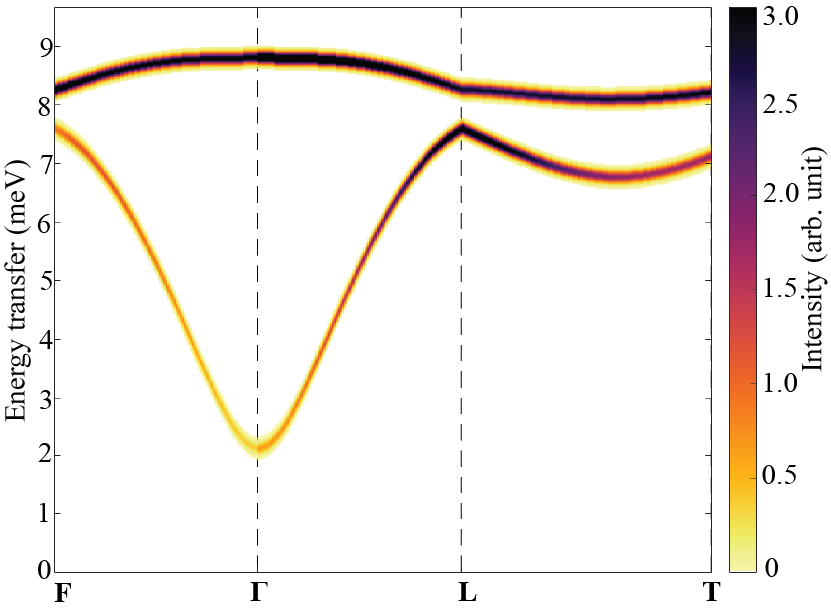}
\caption{Spin wave spectra of \sro~obtained within linear spin wave theory considering
magnetic interactions of Table~\ref{tab:gtable}.}
  \label{fig:fig05}
\end{figure}

We further use the magnetic interactions listed in Table~\ref{tab:gtable} in linear spin-wave theory to obtain the
spin-wave spectra using SpinW package~\cite{spinw}.  
Obtained spectra along various reciprocal space directions is shown
in Fig.~\ref{fig:fig05}.

Several points are to be noted about the spectra. First, one can see that the spectra have gaped along all
directions in reciprocal space with a Goldstone gap of $\sim$ 2 meV. This feature of  
spin-wave spectra may be caused by  
the breakdown of $SU(2)$ symmetry of the isotropic Heisenberg Hamiltonian.
Such a symmetry breaking can be a result of additional Ising like
 Kitaev terms and/or diagonal/off-diagonal anisotropic terms like $\zeta$, $\eta$, $\eta'$
 and $\eta''$. Second, one branch $\sim$ 8 meV in the spectra appears to be dispersion-less.
 It is separated from the dispersing branch by $\sim$ 0.5 meV. Such a feature has previously been observed from the
 inelastic neutron scattering experiments on some of the cobaltates~\cite{co3}, pertinent material candidates for Kitaev
 physics~\cite{co_skp}.
Third, it can observe that the spin-wave spectra near ${\bf \Gamma}$ point is quadratic in nature.
This is in contradiction to the expected linear dispersion of spin-wave dispersion for an antiferromagnetic ground state.

\begin{figure}[ht]
\centering
\includegraphics[width=8.0 cm]{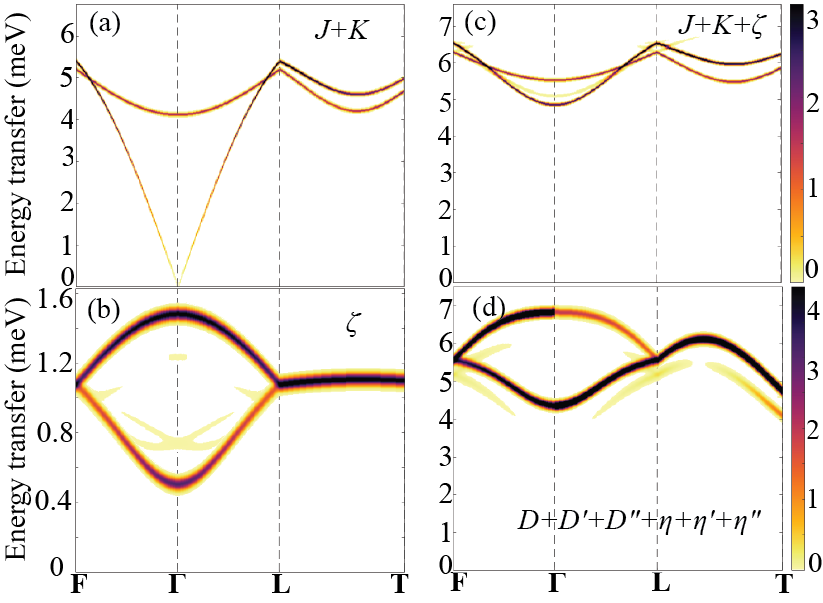}
\caption{Breakdown of spin-wave spectra shown in Fig.~\ref{fig:fig03} to 
individual contributions of combination of various magnetic interactions.
Spectra from, (a) $J + K$ terms, (b) only diagonal anisotropic term $\zeta$, 
(c) $J + K + \zeta$ terms and (d) $D + D' + D" + \eta + \eta' + \eta''$ terms
from Table~\ref{tab:gtable}.}
  \label{fig:fig06}
\end{figure}

In order to investigate the origin of previously mentioned features of spin-wave spectra of \sro, we
 break it down to the contribution of either individual or a specific combination of magnetic interactions and the plots are shown in Fig.~\ref{fig:fig06}.
 Such an analysis can provide useful insights as has been shown in 
 Ntallis {\it et. al.}~\cite{nso_sw} for the case of NaOsO$_3$.
 
 Considering $J$ and $K$ terms together, we immediately obtain both branches with a lower branch, at ${\bf \Gamma}$,
 showing the linear dispersion
  behavior of an antiferromagnet. The plot is shown in  Fig.~\ref{fig:fig06}(a).
However, the spectra are barely gaped in this case due to  
 dominant $J$ over $K$ and which is also responsible for the dispersion width of $\sim$ 5.5 meV of the lower branch.
 Consideration of $K$-only term in the Hamiltonian produces a completely flat branch at $\sim$ 4 meV  (not shown)
 consistent with the previous theoretical
study on Kitev model~\cite{kitaev_th_magnon}.
$\zeta$-only term indeed causes the gap opening along with deviation  
towards a quadratic dispersion at ${\bf \Gamma}$
of the lower branch and as shown in Fig.~\ref{fig:fig06}(b).
However, the energy scale, in this case, is smaller than that of the original spectra in Fig.~\ref{fig:fig05}.
A combination of
$J + K + \zeta$ (Fig.~\ref{fig:fig06}(c)) reproduces some of the features in the more or less similar spectral windows as that of
the original spectra. However, the dispersion width and nature of the lower branch, in this case, are inconsistent with the original
one in Fig~\ref{fig:fig05}. Additionally, near ${\bf \Gamma}$, dispersion of the lower branch appears to be
further deviating from quadratic to higher powers of $\bm{k}$.
The terms $D + D' + D'' + \eta + \eta' + \eta''$ produces similar but relatively flatter branches than the $\zeta$ terms and is shown
 Fig.~\ref{fig:fig06}(d). The spectral energy window, in this case, is similar to that of $J + K + \zeta$ term. Thus one can say
 conclusively that the dominant off-diagonal terms are mainly responsible for the gap in spin-wave spectra of \sro~while the diagonal anisotropic
 term decides the nature of dispersion near ${\bf \Gamma}$ point
 in spin-wave spectra of \sro. The overall spectra which resemble a typical magnetic system with strong frustration
 is a joint effort of all the terms of magnetic Hamiltonian.

\section{Conclusion}

In the quest for new Kitaev candidates,  in this work, we have investigated the electronic and magnetic
properties of \sro.
Through $ab$ $initio$ calculations and a TB model, we show the lowering of cubic symmetry of Rh-O$_6$
octahedra due to additional
trigonal-like distortions which are in contradiction to the previous experimental proposal.
Using the exact diagonalization technique, we show that despite such a distortion, electronic and 
magnetic properties 
of \sro~can be well described
with the pseudo-spin 1/2 framework. The magnetic interactions between these  pseudo-spins were found 
to be highly bond-dependent
anisotropic in nature. We found two particularly noticeable features of the 1NN magnetic 
interactions in \sro~which are, appearance of 
antiferromagnetic Kitaev term and DMI. This may place \sro~in a distinct class of materials as 
previously proposed Kitaev candidates 
shown to have ferromagnetic Kitaev couplings and DMI appears on the 2nd neighbor bonds~\cite{rucl3_ref}. 
The analysis of spin-wave spectra obtained using linear spin-wave theory considering these interactions reveals
the crucial role of diagonal and off-diagonal magnetic interactions in producing a gaped spectrum of \sro. Our theoretical study provides deeper insights about the coupling among 
structural, electronic and magnetic  
degrees of freedom in these compounds and calls for further experimental investigations.

\section{Acknowledgment}
We have greatly benefited from stimulating discussions with Dr. Stephen M Winter and gratefully 
acknowledge his critical reading of our manuscript and valuable feedback. We acknowledge the financial support from the National Key R \& D Program of China 
(Grant No.2018YFA0305601), National 
Natural Science Foundation of China (Grant No. 11725415), and Innovation Program for Quantum Science and Technology 
(Project No. 2021ZD0302600).
This work is also supported partially by the China Postdoctoral Science Foundation
 (No.2022M710231) awarded to Q. Gu.

\appendix

\renewcommand{\thefigure}{A-\arabic{figure}}
\setcounter{figure}{0}

\renewcommand{\theequation}{A-\arabic{equation}}
\setcounter{equation}{0}

\setcounter{table}{0}
\renewcommand{\thetable}{A-\arabic{table}}



\section*{Appendix A: First neighbors Rh-Rh hopping  amplitudes
in \texorpdfstring{\sro}{TEXT} expressed in the basis 
($d^\dagger_{z^2}$, $d^\dagger_{x^2-y^2}$, $d^\dagger_{xz}$, $d^\dagger_{yz}$, $d^\dagger_{xy}$).}
\label{appendix:hop}




\begin{widetext}

\begin{table}[!ht]
\begin{tabular}{ccc}
A-bond & B-bond & C-bond \\
{\tiny
$\begin{pmatrix*}[r]
  -0.0251 & -0.0042 & -0.0015 &  0.0144  &  0.0217 \\
  -0.0042 &  0.0270 &  0.0182 &  0.0131  & -0.0094  \\
  -0.0015 &  0.0182 & -0.0288 &  0.0240  & -0.0319 \\
   0.0144 &  0.0131 &  0.0240 & -0.0369  &  0.0263 \\
   0.0217 & -0.0094 & -0.0319 &  0.0263  & -0.0199  
 \end{pmatrix*}$
}
&
{\tiny
$\begin{pmatrix*}[r]
  0.0245 &  0.0006 & -0.0302 & 0.0517 &  0.0141 \\
  0.0368 & -0.0040 & -0.0626 & 0.0285 &  0.0339  \\
 -0.0334 & -0.0077 & -0.0100 & 0.0009 & -0.0192 \\
 -0.0791 & -0.0004 &  0.0633 & 0.0175 &  0.0087 \\
  0.0476 & -0.0294 &  0.0256 & 0.0130 &  0.0033 
 \end{pmatrix*}$
 }
 &
 {\tiny
$\begin{pmatrix*}[r]
-0.0083  &  0.0134  & -0.0001 & -0.0217 &  -0.0419  \\
  0.0134 &  -0.0245 &  0.0618 &  0.0233 &  0.0622  \\
 -0.0001 &   0.0618 & -0.0173 & -0.0057 &  0.0128 \\
 -0.0217 &   0.0233 & -0.0057 & -0.0194 &  0.0103 \\
 -0.0419 &   0.0622 &  0.0128 &  0.0103 & -0.0684 \\
 \end{pmatrix*}$ 
 }
 \\
\end{tabular}
\caption{First neighbor Rh-Rh hopping amplitudes on different types of bonds shown in Fig.~\ref{fig:fig01}(c)}.
\end{table}
\end{widetext}

\end{document}